\documentclass[aps,prl,twocolumn,showpacs,preprintnumbers,amsmath,amssymb]{revtex4}

\usepackage{graphicx}
\usepackage{dcolumn}
\usepackage{bm}

\begin{document}

\title{Counting atoms using interaction blockade\\in an optical superlattice}

\author{P. Cheinet,$^{1}$ S. Trotzky,$^{1}$ M. Feld,$^{1,2}$ U. Schnorrberger,$^{1}$\\M. Moreno-Cardoner,$^{1}$ S. F\"olling,$^{1}$ I. Bloch$^{1}$}
\altaffiliation[Corresponding author: ]{bloch@uni-mainz.de.}
\affiliation{$^{1}$Institut f\"ur Physik, Johannes Gutenberg-Universit\"at, 55099 Mainz, Germany\\$^{2}$Fachbereich Physik, Technische Universit\"at Kaiserslautern, 67663 Kaiserslautern, Germany}

\date{\today}

\begin{abstract}

We report on the observation of an interaction blockade effect for ultracold atoms in optical lattices, analogous to Coulomb blockade observed in mesoscopic solid state systems. When the lattice sites are converted into biased double wells, we detect a discrete set of steps in the well population for increasing bias potentials. These correspond to tunneling resonances where the atom number on each side of the barrier changes one by one. This allows us to count and control the number of atoms within a given well. By evaluating the amplitude of the different plateaus, we can fully determine the number distribution of the atoms in the lattice, which we demonstrate for the case of a superfluid and Mott insulating regime of $^{87}$Rb.
\end{abstract}

\maketitle


Interactions between individual particles can drastically alter the ground state properties of a many-body system and are known to crucially determine the transport properties in mesoscopic solid state systems. For instance, in quantum dots, Coulomb blockade \cite{Lambe69,Averin86,Meirav89,Reimann02} can prevent current flow at small bias voltages and induces tunneling resonances involving a single electron crossing the dot. It demonstrates the quantization of charge in a striking manner and enables counting and control of the charge carriers in the dot with single electron resolution, provided that the thermal energy is lower than the interaction energy. For the case of ultracold atoms, similar interaction blockade effects have been predicted \cite{Micheli:2004,Capelle07,Seaman:2007,Dounas-Frazer:2007,Lee:2008} that could furthermore enable the generation of non-classical matter wave states or the preparation of highly correlated and entangled samples of atoms. Here we present a direct measurement of such an interaction blockade with atoms in a bichromatic optical superlattice \cite{Ritt06,Sebby06,Foelling07}. By adiabatically changing each site of a tetragonal lattice into a biased double well, the atoms experience an interaction blockade forcing them to distribute over the two sides of the double wells in a defined manner. Such an ``interacting beam splitter" (IBS) for atoms can be used to obtain ``constructed pairs" of atoms \cite{Sebby07} when slowly recombining a double well, or to avoid creating unwanted Fock states when splitting \cite{Trotzky07}. Observing the interaction blockade--induced tunneling resonances with an intra double well site sensitive detection \cite{Sebby07,Foelling07} provides a novel method of measuring the number distribution in the lattice \cite{Greiner:2002b,Gerbier06,Campbell06,Sebby07}. We demonstrate this by fully determining the atom number distribution of an ultracold degenerate cloud of $^{87}$Rb in the superfluid or strongly interacting regime of a Mott insulator \cite{Mott}.

We consider the transport of ultracold atoms between the two sides of a double well generated with an optical superlattice combined with two transverse lattices \cite{Foelling07}. The superlattice potential is given by:
\begin{equation}\label{eq:Pot}
V(x) = V_s \cos^2(4 \pi x/\lambda_l -\phi) + V_l \cos^2(2 \pi x/\lambda_l)\,,
\end{equation}

where $\phi$ is the relative phase between the short and long period lattices. $V_{s,l,t}$ denote the lattice depths of the short, long and transverse lattices, expressed in units of their corresponding recoil energies $E^{s,l,t}_{r}=h^2/2 M \lambda^{2}_{s,l,t}$. The optical wavelengths are $\lambda_{{s,l,t}}$ and $M$ is the atomic mass. We consider isolated double wells, which we describe with a two-mode Bose-Hubbard type Hamiltonian:
\begin{eqnarray}\label{eq:Hamiltonian}
\hat H
  = -J\left(\hat a^\dagger_{L} \hat a_{R}
                +\hat a^\dagger_{R} \hat a_{L}\right)
                -\frac{\Delta}{2}(\hat n_{L}-\hat n_{R})\nonumber\\
+\frac{U}{2}\left(\hat n_{L} \left(\hat n_{L}-1\right)
                + \hat n_{R} \left(\hat n_{R}-1\right)\right),
\end{eqnarray}

where the operators $\hat a^\dagger_{L,R}$ and $\hat a_{L,R}$ create and annihilate an atom in the left and right well, $\hat n_{L,R}$ count the atom number per well, $J$ is the tunneling matrix element, $\Delta$ the potential bias along the double well axis and $U$ the onsite interaction energy between two atoms. The state of the system can be described in the basis of the Fock states $|n_{L},n_{R}\rangle$, where $n_{L,R}$ represent the discrete atom numbers on the left and right hand side of the double well with a total atom number $n=n_{L}+n_{R}$.

\begin{figure}
\includegraphics[width=.8\columnwidth]{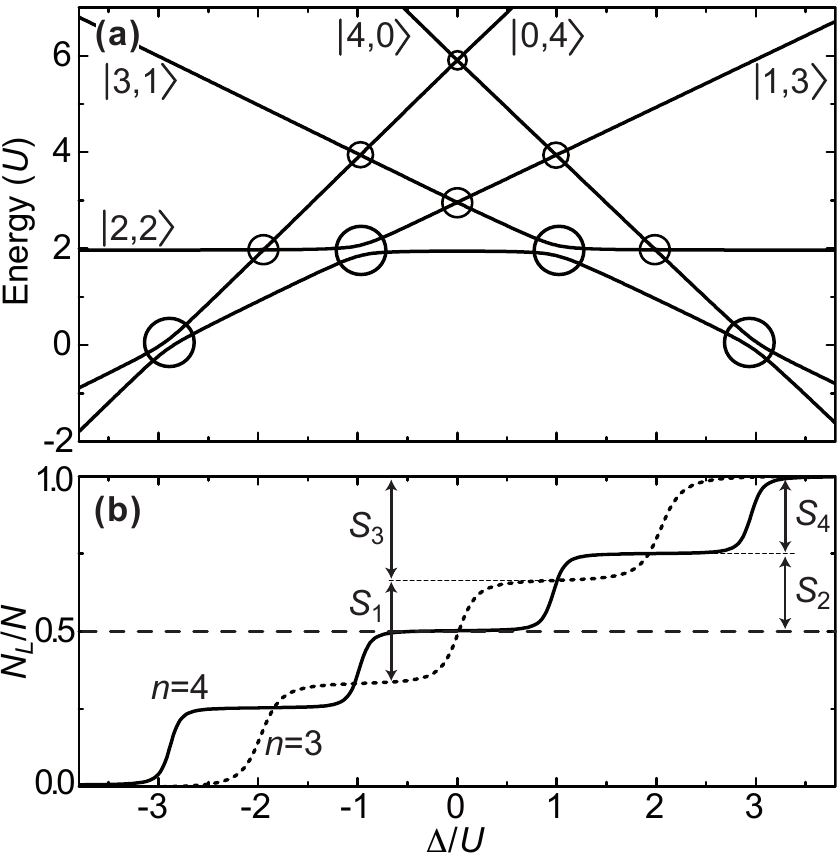}
\caption{\label{fig:Energies} (a) Eigenenergies in a double well filled with 4 atoms versus $\Delta$ for $U/J = 30$. The single particle resonances (large circles) take place between the two energetically lowest states. Higher order tunneling resonances occur with smaller couplings (smaller circles). Away from these resonances, the eigenstates are Fock states. (b) Left-well normalized population versus $\Delta$ for the ground state with 4 atoms (solid line) and 3 atoms (dashed line).}
\end{figure}

Let us focus on the blockade and calculate for the ground state with $n$ atoms $| \Psi_g\rangle_n$ the probability of each atom to be in the left well: $\langle \hat n_{L} \rangle_n /n = \,_n\!\langle\Psi_g|\hat n_{L}| \Psi_g\rangle_n/n$. For non-interacting particles, all Fock states are degenerate at $\Delta=0$ and all atoms would gather in the lower energy well for $|\Delta| \gg J$. For $U \neq 0$, we diagonalize the Hamiltonian (\ref{eq:Hamiltonian}) to obtain the eigenstates and corresponding eigenenergies of the system. For instance, the eigenenergies for $n=4$ are plotted against the bias $\Delta$ in Fig.~\ref{fig:Energies}a for $U/J=30$. At small bias $|\Delta|<U$, the ground state of the system mainly consists of $|2,2\rangle$, and the transport of an additional atom to the lower well is blocked. When $\Delta = U$, a tunneling resonance occurs between the two states $|2,2\rangle$ and $|3,1\rangle$, leaving the latter as the ground state for larger bias until the next resonance is reached for $\Delta=3U$. 

In the experiment, several thousands of double wells are populated simultaneously and our measurements can only reveal the overall number of atoms in the left (right) wells $N_{L}$ ($N_{R}$) and the total atom number $N=N_{L}+N_{R}$ in the lattice. The ratio $N_{L}/N$ is given by the statistical average: $N_{L}/N=\sum_n g_n \langle \hat n_L \rangle_n/n$, where $g_n$ denotes the fraction of atoms located in sites with occupation $n$. For a homogeneous system with an integer filling $n$, $N_L/N$ equals the quantum mechanical expectation value for a single double well. Such a case is displayed in Fig.~\ref{fig:Energies}b for $n=3$ and $n=4$. When the magnitude of the bias potential is increased, one observes a succession of $n$ steps with an amplitude of $1/n$, each corresponding to one atom changing sides within the double well. With $i=1$ corresponding to $\Delta =0$, we define $S_i=\left ( N_{L}[\Delta=(i-1/2)U]-N_{L}[\Delta=(i-3/2)U]\right)/N$ to be the $i$-th step amplitude, occurring at $\Delta=(i-1)U$. Even (odd) fillings exhibit only even (odd) steps and the step heights for given $g_n$ are $S_i=\sum_{j=0}^{\infty}g_{i+2j}/(i+2j)$. More importantly, the distribution function $g_n$ can be fully derived from the measured step heights $S_i$ as:
\begin{equation}\label{eq:NumbDist}
g_n = n \left(S_n-S_{n+2}\right).
\end{equation}


Our experimental sequence begins by loading a $^{87}$Rb Bose-Einstein condensate (BEC) into a tetragonal optical lattice, with periodicities of $\lambda_{{l}}/2=765$\,nm on one axis and $\lambda_t/2=420$\,nm on the transverse axes. During this loading period the lattices are first ramped to $V_l=2\,E^{{l}}_{r}$ and $V_t=6\,E^{{t}}_{r}$ in $100$\,ms, with all ramps optimized for an adiabatic loading \cite{Gericke07}. A second ramp over $200$\,ms to final depths of $V_l=40\, E^{{l}}_{r}$ and $V_t=50\,E^{{t}}_{r}$ brings the system into a Mott insulating regime. All three tunneling frequencies are kept equal during the second ramp, which ensures a redistribution of the density on all axes on the same time scale.  

\begin{figure}
\includegraphics[width=.8\columnwidth]{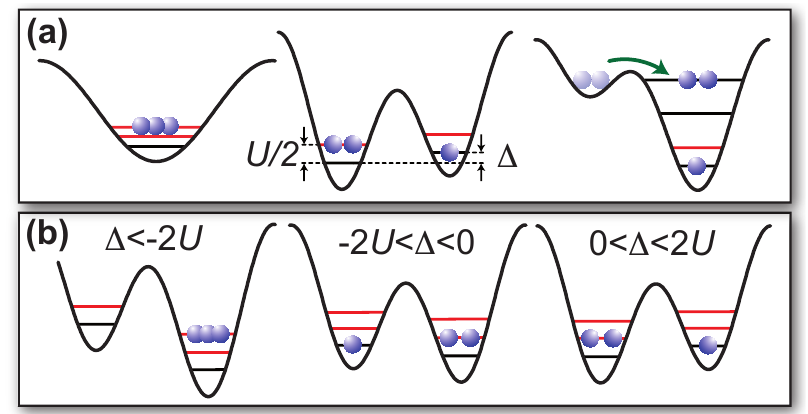}
\caption{\label{fig:Sequence} (a) Experimental sequence. After loading the BEC into a 3D lattice, each site is split slowly into a biased double well. The atoms in the left well are transferred into a high vibrational level to be counted in a separate Brillouin zone. (b) The atoms distribute over the two wells depending on the bias $\Delta$ and the interaction energy $U$. The single particle vibrational states are denoted by black lines, whereas states shifted by interaction energies $1/2\,U$, $U$ etc. per particle are shown in red (energy separations not to scale).}
\end{figure}

The interaction blockade is observed after subsequently ramping up the short lattice to $V_{{s}}=44\, E^{{s}}_{r}$ in $20$\,ms for different superlattice phases $\phi$, i.e. different potential biases $\Delta$, as depicted in Fig.~\ref{fig:Sequence}a. The rising barrier realizes an IBS, which forces the atoms to redistribute via tunneling (Fig.~\ref{fig:Sequence}b) over the two newly-formed wells.

\begin{figure}
\includegraphics[width=.8\columnwidth]{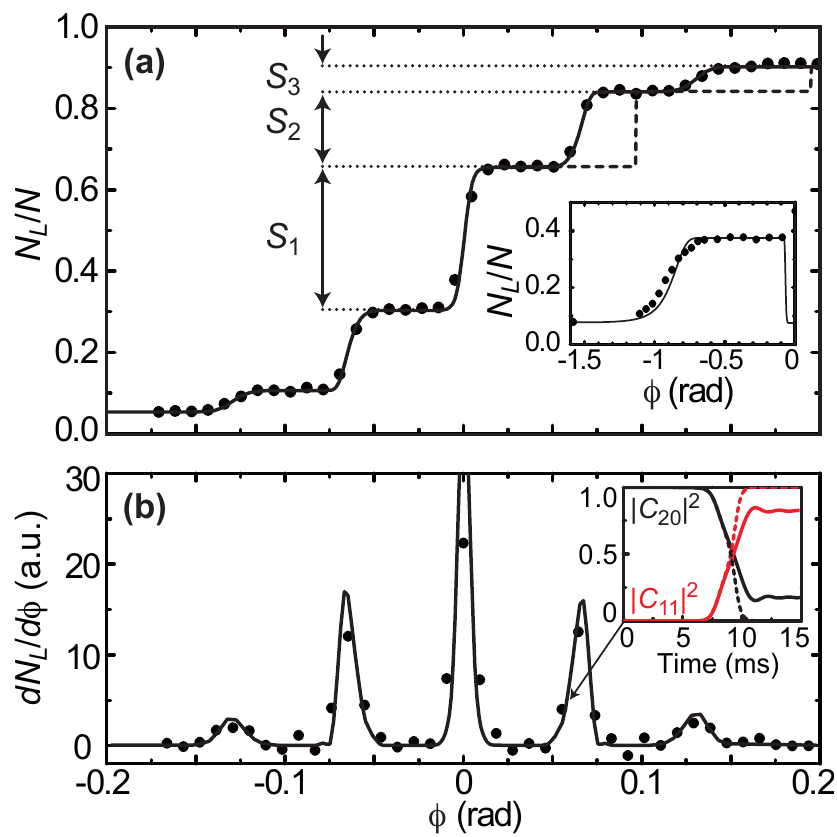}
\caption{\label{fig:Steps} (a) Left well population, measured as a function of the phase $\phi$ between the short and long lattices (points), theory assuming full adiabaticity during the IBS (dashed line) and theory accounting for the dynamical evolution during the splitting (solid line). For both theory curves, the amplitudes of the steps were extracted from the experimental data. Inset: tunneling resonance to higher vibrational levels. (b) Tunneling resonances, shown in the derivative of the left well population. Inset: Fock state coefficients in the SPE basis during the time evolution (solid) with $\phi=0.06$\,rad, compared to the eigenstates at the corresponding time (dashed).}
\end{figure}

In order to read out the number of atoms in the left well after the IBS sequence, we carry out the detection sequence shown in Fig.~\ref{fig:Sequence}a (right panel). By increasing the bias in $20$\,ms to a fixed phase of $\phi_0=-1.1$\,rad, a highly tilted double well is formed. Turning off the short lattice exponentially with a time constant of $500$\,$\mu$s subsequently transfers the lifted atoms into a high vibrational level of the remaining single well \cite{Sebby07}. Finally, a band mapping technique \cite{Kastberg95,Foelling07} allows us to count the populations in different energy bands by counting the population of the different Brillouin zones after time-of-flight, which directly gives $N_{L}$ and $N_{R}$. The resulting occupations $N_{L} / N$ as a function of the superlattice phase $\phi$ are displayed in Fig.~\ref{fig:Steps}a for $N \approx 2\times10^{5}$. We observe pronounced plateaus separated by tunneling resonance steps up to $S_3$. For very large bias, one would expect all atoms to occupy the lower energy well, however in our measurements about 5-10\% of the atoms are still detected in the Brillouin zone corresponding to the side of the double well with higher potential energy. We suspect this to be a result of excitations created during the fast transfer in the detection step, heating due to spontaneous photon scattering in the lattice or collisions during the time-of-flight period which can mix the measured populations between the Brillouin zones. 

In addition to the resonances induced by the interaction blockade, tunneling resonances to higher excited vibrational levels may be observed for very large applied potential biases when the system is not in its ground  state \cite{Bharucha:1997,Sias:2007}. In order to reveal their presence, we optimize the loading to mostly obtain double wells with two atoms in the $|1,1\rangle$ Fock state. The left atom is then transferred into the first vibrational band of the long lattice. The IBS and detection sequence are applied again, and we obtain the curve displayed in the inset of Fig.~\ref{fig:Steps}a which now shows a broad tunneling resonance for a bias of $\Delta\approx \hbar \omega_{ho}$ as expected from theory, when adding the second‚ vibrational level to the Fock state basis (solid).

For the lowest band, we plot the steps expected for our lattice configuration and assuming full adiabaticity during the IBS in Fig.~3a (dashed line). The theoretical result has been obtained from the single-particle eigenstates (SPE) of the biased double wells, and using $a_{s}=5.31$\,nm for the scattering length \cite{Kempen02}. The experimental and theoretical results clearly show a discrepancy in the position and width of the steps, which we attribute to a loss of adiabaticity during the splitting process at moderately deep lattices. To confirm this, we have performed a theoretical analysis of the time evolution of the splitting process using the basis set of the two lowest SPEs. We denote the population of these two states  by $|n_{g},n-n_{g}\rangle_{\rm SPE}$, where $n_g$ is the population of the lower energy state. The general state of the $n$ particles can then be written as $| \Psi (t) \rangle=\sum_{n_{g} =0}^{n} C_{n_{g},n-n_{g}} (t) |n_{g},n-n_{g}\rangle_{\rm SPE}$. We determine the time evolution of the relevant amplitudes by a numerical solution of the Schr\"odinger equation for the ramps used in the experiment. The resulting theoretical curve is displayed in Fig.~\ref{fig:Steps} (solid line) and the inset of Fig.~3b shows the time evolution of the coefficients $C_{2,0}$ (black) and $C_{1,1}$ (red) for a superlattice phase of $\phi=0.06$\,rad close to the $S_2$ resonance. We see that the system is not able to completely follow the ground state during the splitting process. Both position and width of the resonances observed in the experiment are in good agreement with the calculation (see Fig.~\ref{fig:Steps} (solid line)) when the interaction strength is reduced by 20\% from the one obtained for the SPE's \cite{note}. 
We note, however, that although the positions of the resonance steps are shifted to smaller biases due to this non-adiabatic evolution, the amplitude of the steps is fully preserved. The relevant atom number distribution is simply frozen out at an intermediate short-lattice depth of $\approx 20\,E_r^s$ during the ramp up.

\begin{figure}
\includegraphics[width=.8\columnwidth]{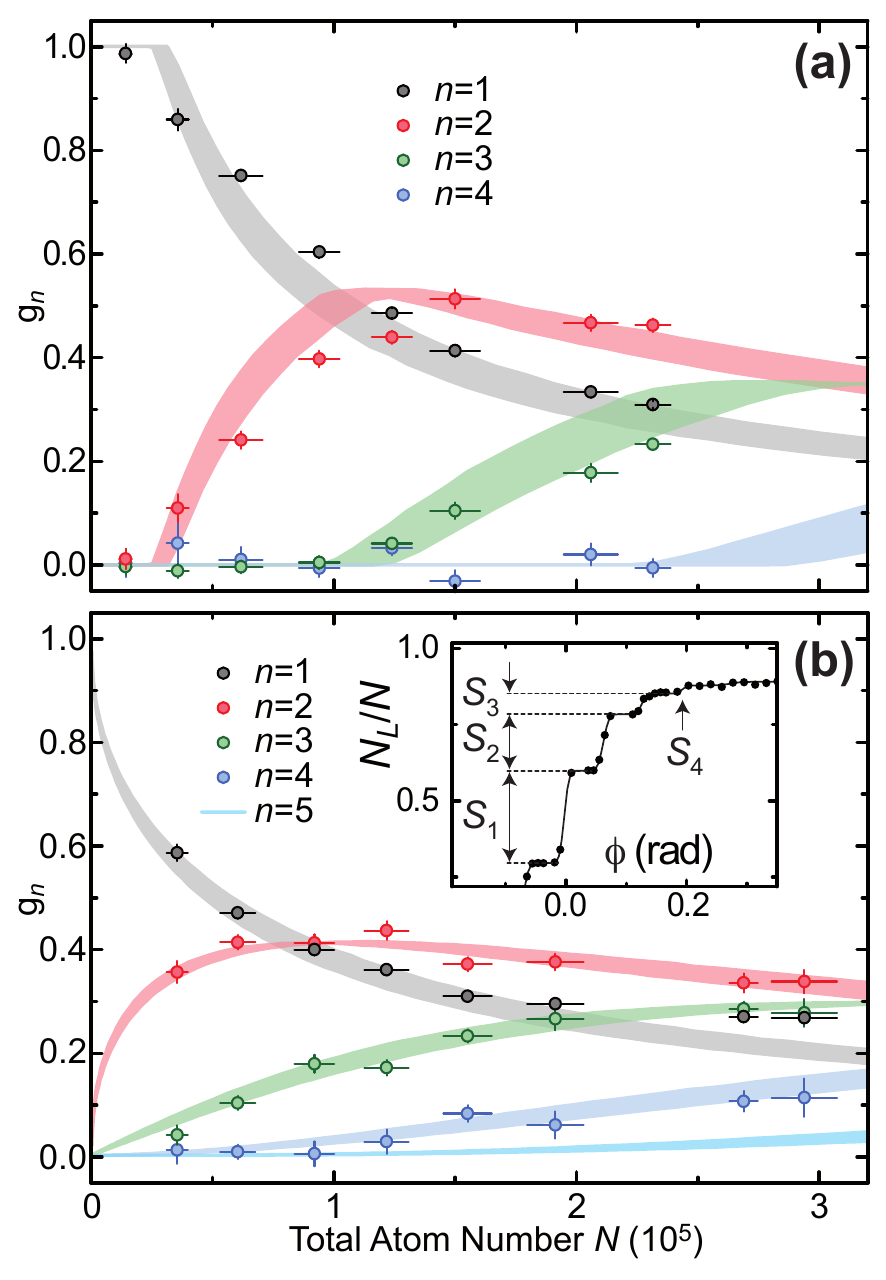}
\caption{\label{fig:NumberStat} Number distribution versus the total atom number, measured in the Mott insulator regime (a) and in the superfluid  regime (b). The shaded lines are predictions for the experimental parameters (see text) and $T=0$. Their widths represent the uncertainties in atom number and lattice depth calibrations. The error bars reflect the standard deviation of the measurements. The inset shows a typical occupation measurement in the superfluid regime for $N\approx2\times10^5$ with four visible steps.}
\end{figure}
As outlined in the introduction, the resonance amplitudes $S_i$ can be used to determine the number distribution $g_n$ in the initial tetragonal lattice. We have carried out such an analysis both for the case of a Mott insulator and a superfluid as a function of the total atom number. We take up to $3$ measurements in the center of each plateau to obtain the resonance amplitudes $S_i$ with a typical standard deviation of $\leq 0.5\%$. In the analysis, we have truncated the distribution above a maximum filling of $n=4$ and rescaled the measured amplitudes by an overall factor, such that $S_1+\sum_{n=2}^4 2 S_n=1$. 

The measured number distribution is displayed in Fig.~\ref{fig:NumberStat}a in the Mott insulator regime, for which we use the previous loading sequence. Within our uncertainties, we detect only an $n=1$ component for small $N$, which corresponds to a single Mott insulator shell. For increasing total atom numbers, the next shells appear at $N \approx 3\times10^4$ for $n=2$ and $N \approx 1\times10^5$ for $n=3$. We compare our results with a $T=0$ and $J=0$ calculation of the number distribution for our experimental parameters \cite{note2} and a lattice depth of $V_s=18\,E_r^s$ and $V_t=24 E_r^t$, which is the configuration where we expect the density distribution to be frozen during the ramp up. The shaded areas in Fig.~\ref{fig:NumberStat} correspond to an uncertainty of $\pm10\%$ in the atom number calibration and $\pm2\%$ in the lattice depths and waists. 

In order to measure the number distribution in the superfluid regime, we ramp up the long and transverse lattices to $V_{{l}}=6\,E^{{l}}_{r}$ and $V_{{t}}=6\,E^{{t}}_{r}$ in a first ramp over $200$\,ms. After a hold time of $40$\,ms, we freeze out the number distribution by ramping the lattices to $V_{{l}}=40\,E^{{l}}_{r}$ and $V_{{t}}=50\,E^{{t}}_{r}$ in $1$\,ms. The inset in Fig.~\ref{fig:NumberStat}b shows a left-well occupation measurement realized in the superfluid regime with $N \approx 2\times10^{5}$ atoms. Up to the fourth resonance step can now be identified. We observe that even for the smallest atom numbers, the superfluid contains visible fractions of up to $n=3$ fillings. A simple model which assumes both $T=0$ and a Thomas-Fermi profile with a Poissonian onsite number distribution overestimates the $n=3$ and $n=4$ fractions. Therefore we consider the Gutzwiller approximation adapted for bosons \cite{Rokhsar91}. It uses a modified Poissonian distribution $s^{n(n-1)}\lambda^n / n!$, which accounts for a possible interaction induced number squeezing through a squeezing parameter $s=1/(1+\frac{U}{2zJ})$ where $z$ is the number of nearest neighbors. For anisotropic lattices, the term $zJ$ is replaced by the sum over all nearest-neighbor couplings. We find that the Gutzwiller distribution models our measurement quantitatively for a squeezing parameter $s=0.66(5)$ (see Fig.~\ref{fig:NumberStat}b shaded areas). The squeezing beyond the expected value of $s=0.84$ could be due to the finite ramp time during the freeze out, or due to stronger correlation effects that are not captured by the Gutzwiller approximation.

In conclusion, we have shown how interaction blockade can be used to count and control the population of ultracold atoms on a single side of a double well, both with single atom resolution. Furthermore, we have demonstrated the ability to  measure the full number statistics in a 3D optical lattice for different quantum phases using this blockade effect. By using our method, it is also possible to check predicted phase diagrams in superlattices, in particular to verify the existence of fractional filling loopholes \cite{BuonsanteFrac04}. Furthermore, by exploiting avoided crossings in higher order tunneling terms, one could create several non-classical and entangled number states within the double well in a massively parallel way.

\begin{acknowledgments}
We would like to acknowledge funding of the project through the DFG, the EU (MC-EXT QUASICOMBS, IP SCALA), AFOSR and DARPA (OLE). The authors thank B. Paredes, J. Thompson and F. Gerbier for helpful discussions.
\end{acknowledgments}



\begin{thebibliography}{28}
\expandafter\ifx\csname natexlab\endcsname\relax\def\natexlab#1{#1}\fi
\expandafter\ifx\csname bibnamefont\endcsname\relax
  \def\bibnamefont#1{#1}\fi
\expandafter\ifx\csname bibfnamefont\endcsname\relax
  \def\bibfnamefont#1{#1}\fi
\expandafter\ifx\csname citenamefont\endcsname\relax
  \def\citenamefont#1{#1}\fi
\expandafter\ifx\csname url\endcsname\relax
  \def\url#1{\texttt{#1}}\fi
\expandafter\ifx\csname urlprefix\endcsname\relax\def\urlprefix{URL }\fi
\providecommand{\bibinfo}[2]{#2}
\providecommand{\eprint}[2][]{\url{#2}}

\bibitem{Lambe69} J. Lambe and R.~C. Jaklevic, Phys.\ Rev.\ Lett. \textbf{22}, 1371 (1969). 

\bibitem{Averin86} D.~V. Averin and K. Likharev, J. Low Temp. Phys. \textbf{62}, 345 (1986). 

\bibitem{Meirav89} U. Meirav~{\it et al.}, Phys.\ Rev.\ B \textbf{40}, 5870 (1989). 

\bibitem{Reimann02} S.~M. Reimann, Rev.\ Mod.\ Phys. \textbf{74}, 1283 (2002).

\bibitem{Micheli:2004} A. Micheli~{\it et al.}, Phys.\ Rev.\ Lett. \textbf{93}, 140408 (2004). 

\bibitem{Capelle07} K. Capelle~{\it et al.}, Phys.\ Rev.\ Lett. \textbf{99}, 010402 (2007). 

\bibitem{Seaman:2007} B.~T. Seaman~{\it et al.}, Phys.\ Rev.\ A \textbf{75}, 023615 (2007).

\bibitem{Dounas-Frazer:2007} D.~R. Dounas-Frazer, A.~M. Hermundstad and L.~D. Carr, Phys. Rev. Lett. \textbf{99}, 200402 (2007).

\bibitem{Lee:2008} C. Lee, L.-B. Fu and Y.~S. Kivshar, Europhys. Lett. \textbf{81}, 60006 (2008).

\bibitem{Ritt06} G. Ritt~{\it et al.}, Phys. Rev. A \textbf{74}, 063622 (2006).

\bibitem{Sebby06} J. Sebby-Strabley~{\it et al.}, Phys. Rev. A \textbf{73}, 033605 (2006).

\bibitem{Foelling07} S. F\"{o}lling~{\it et al.}, Nature \textbf{448}, 1029 (2007).

\bibitem{Sebby07} J. Sebby-Strabley~{\it et al.}, Phys.\ Rev.\ Lett. \textbf{98}, 200405 (2007).

\bibitem{Trotzky07} S. Trotzky~{\it et al.}, Science \textbf{319}, 295 (2008).

\bibitem{Greiner:2002b} M. Greiner~{\it et al.}, Nature \textbf{419}, 51 (2002). 

\bibitem{Gerbier06} F. Gerbier~{\it et al.}, Phys. Rev. Lett. \textbf{96}, 0932010401 (2006). 

\bibitem{Campbell06} G.~K. Campbell~{\it et al.}, Science \textbf{313}, 649 (2006). 
  
\bibitem{Mott} M.~P.~A. Fisher~{\it et al.}, Phys.\ Rev.\ B \textbf{40}, 546 (1989); D. Jaksch~{\it et al.}, Phys.\ Rev.\ Lett. \textbf{81}, 3108 (1998); M. Greiner~{\it et al.}, Nature \textbf{415}, 39 (2002); T. St\"{o}ferle~{\it et al.}, Phys.\ Rev.\ Lett. \textbf{92}, 130403 (2004); I.~B. Spielman, W.~D. Phillips and J.~V. Porto, Phys.\ Rev.\ Lett. \textbf{98}, 080404 (2007).

\bibitem{Gericke07} T. Gericke~{\it et al.}, J. Mod. Opt. \textbf{54}, 735 (2007).

\bibitem{Kastberg95} A. Kastberg~{\it et al.}, Phys. Rev. Lett. \textbf{74}, 1542 (1995).

\bibitem{Bharucha:1997} C.~F. Bharucha~{\it et al.}, Phys. Rev. A \textbf{55}, R857 (1997). 

\bibitem{Sias:2007} C. Sias~{\it et al.}, Phys. Rev. Lett. \textbf{98}, 120403 (2007).

\bibitem{Kempen02} E.~G.~M. van Kempen~{\it et al.}, Phys.\ Rev.\ Lett. \textbf{88}, 093201 (2002).

\bibitem{note} A variational analysis yields a reduction of $U$ by approx. 10\% due to a broadening of the onsite wavefunction for higher occupations. 

\bibitem{note2} Beam waists (1/$e^2$) of $\approx 140$\,$\mu$m, and magnetic trapping frequencies of $\omega_{\rm magn} \approx 2\pi \times 15$\,Hz.

\bibitem{Rokhsar91} D.~S. Rokhsar and B.~G. Kotliar, Phys. Rev. B \textbf{44}, 10328 (1991).

\bibitem{BuonsanteFrac04} P. Buonsante, V. Penna and A.\,Vezzani, Phys.\ Rev.\ A \textbf{70}, 061603  (2004).
  
\end{thebibliography}
\end{document}